\documentclass[preprint,preprintnumbers,amsmath,amssymb]{revtex4}
\usepackage{graphicx} % include figure files
\usepackage{dcolumn} % align table columns on decimal point
\usepackage{bm} % bold math
\usepackage{natbib}
\usepackage{hyperref}
\usepackage{dcolumn}% Align table columns on decimal point
\hypersetup{colorlinks=true,linkcolor=blue,citecolor=blue,urlcolor=blue}
\begin{document}

\title{Double Gated Single Molecular Transistor for Charge Detection}

\author{S. J. Ray}
\email{ray.sjr@gmail.com}
\affiliation{CEA, INAC-SPSMS, UMR-E CEA/UJF-Grenoble 1, 17 Rue Des Martyrs, F-38054 Grenoble Cedex 9, France}

\author{R. Chowdhury}
\affiliation{Department of Civil Engineering, Indian Institute of Technology Roorkee, Roorkee 247 667, India}

%\date{\today}
\begin{abstract}
The electrostatic behaviour of an 1,3-Cyclobutadiene (C$_{4}$H$_{4}$) based Single Molecular Transistor (SMT) has been investigated using the first principle calculation based on Density functional Theory and non-equilibrium Green's function approach. While the molecule is placed on top of a dielectric layer (backed by a metallic gate) and weakly coupled between the Source/Drain electrodes, the charge stability diagram revealed the presence of individual charge states in the Coulomb Blockade regime. This gets affected significantly on addition of an another gate electrode placed on the top of the molecule. This modified double-gated geometry allows additional control of the total energy of the system that is sensitive to the individual charge states of the molecule which can be used as a charge sensing technique operational at room temperature.
\end{abstract}



\maketitle
\setcounter{figure}{0}
\setcounter{table}{0}

\section{Introduction}

In the area of nanoelectronics, single electron transistor (SET) is well known due to its quantised  nature of transport since its discovery \cite{Devort1992} that has found application for sensitive charge detections \cite{Delsing1998, Berman1997}. The packing density of the MOSFET devices in the present day CMOS technology is limited by the gate lengths accessible using various lithography techniques and the tunnelling issues. As a result of which the industry is constantly looking for useful alternatives in next generation technology to achieve faster switching performance and speed. SET's can be useful in this direction due to its small size and low power operation and performance that can be easily integrated towards large scale fabrication in future. While the conventional metallic SET's were made using a metallic island artificially placed between the S/D electrodes, in the recent times growing interest was found in replacing the metallic island by an organic molecule and hence the area of molecular electronics was developed that has made significant progress towards developing single molecular devices (SMT) with the organic molecules as the active component of the circuit \cite{Reed1997, Heath2003}. Operation of such devices have been in presence for the past years which was first demonstrated by Reed $et.~al$ \cite{Reed1997} using Benzene-1,4-dithiolate as the active molecule. Experimental realisations of such devices were made successfully in Carbon based systems \cite{Dekker1998, Dekker2001}, Benzene \cite{Stokbro2010}, Oligophenylenevinylene (OPV) \cite{Kubatkin2003, Osorio2007}, Fullerene \cite{Park2000}, Dipyridylamide \cite{Chae2006} and to design logic circuits \cite{Dekker2001Science} etc. However, understanding the detailed nature of electronic transport in such devices in still under study and has not been well understood in details. In the strong coupling (SC) limit, the conduction behaviour is usually dominated by the coherent electron transport that often overestimates the current/conductance and in the weak coupling (WC) limit, the sequential nature of transport is governed by the orthodox theory of Coulomb Blockade that underestimates the value of the energy gap \cite{Kasper2009}. While the transport behaviour is usually estimated semi-empirically using a combination of Non-equilibrium Green's function formalism (NEGF) and Density Functional Theory (DFT)  \cite{Brandbyge2002} in the SC limit, the incoherent transport behaviour in the WC limit was formalised by a recent approach proposed by Kaasberg $et.~ al$ \cite{Kaasbjerg2008} and Strokobro $et.~al$ \cite{Stokbro2010} has introduced this within a DFT framework for the estimation of various energy levels. This approach has been used computationally in the recent times to estimate the charging energy of systems based on Benzene \cite{Ventra2000, Stokbro2010}, Fullerene \cite{Stokbro2010}, Napthalene \cite{Parashar2012}, DNA chain \cite{Guo2012} etc. which has found excellent agreement with the experimental results.

Due to the incoherent nature of transport, a SMT is incredibly sensitivity to its charge state similar to a SET. While an incoming charge is capacitively coupled to the SMT molecule, any change in the charge state can be sensed through the charge stability diagram. In a conventional SMT,  the gate electrode is placed in very close proximity of the island (or the molecule) to control the chemical potential independently. Here in this present case, we have modified this geometry to introduce an additional gate electrode on top of the molecule to achieve better electrostatic control over the device and to find its influence on the energy levels of the molecule and to demonstrate its usefulness as a sensitive charge detector.

\section{System description and Computational Recipe}

\begin{figure*}
\begin{center}
\includegraphics[width=17cm]{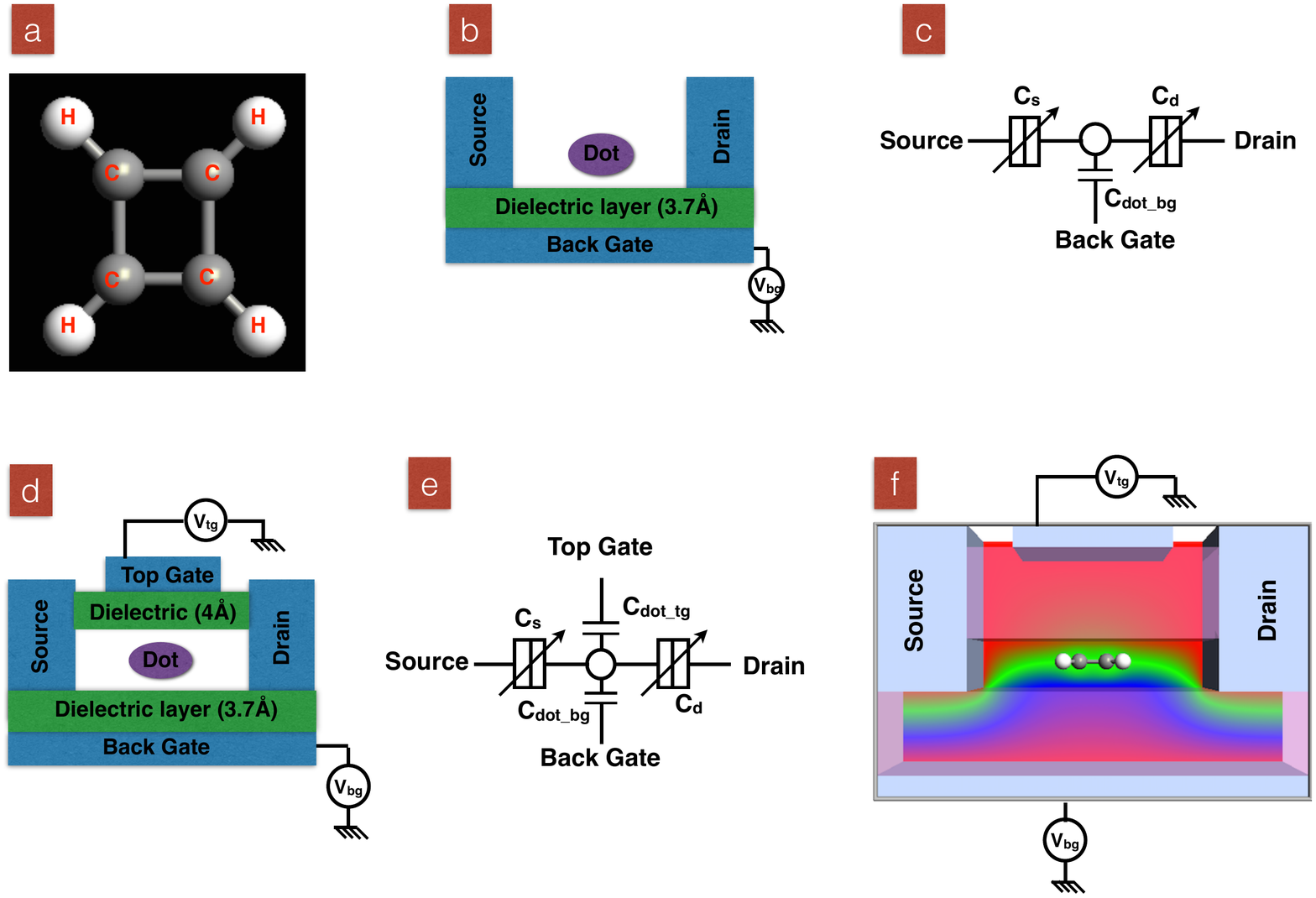}
\caption{{\small (a) Schematic of 1,3-Cyclobutadiene molecule, (b) 2D Schematic of the Single gated SMT under investigation (not upto the scale) with the Cyclobutadiene molecule (referred as the `dot') positioned on top of the dielectric layer and the backgate connected to its back, (c) Junction capacitances corresponding to the device in Fig.~\ref{fig.1}(b), (d) Schematic of the Double gated SMT with an additional gate connected to the top,  (e) Capacitance network for the Double gated SMT and (f) A sample illustration of the distribution of the induced electrostatic potential in different regions of the device for an equal but opposite values of the two gate voltages at $V_{d} = 0$ indicating the molecule stays in an equipotential (green) region.}}
\label{fig.1}
\end{center}
\end{figure*}

In this present work, the active component of the the system under investigation is an 1,3-Cyclobutadiene molecule working as the `island' in the SMT devices. The molecule has a square central structure with 4 Hydrogen atoms placed symmetrically at each corners of the Carbon atoms as illustrated in Fig.~\ref{fig.1}(a). The small and planar structure of Cyclobutadiene is advantageous in such a device as the influence of electrical polarisation is significantly less than the direct gate-molecule coupling that determines the performance of a transistor. For the single and double gated SMT \cite{Ray2014b} as illustrated in Fig.~\ref{fig.1}(b),(d), the molecule was placed symmetrically between the source (S) and drain (D) electrodes on top of a dielectric slab with the molecular plane lying parallel to the dielectric surface. For the single gated device, the dielectric layer is of thickness ($d_{b}$) = 3.7 \AA\, and with a dielectric constant of 10$\varepsilon_{0}$ which is connected to a metallic backgate electrode of thickness 1 \AA\, as illustrated in Fig.~\ref{fig.1}(b). In the case of the double gated device (see Fig.~\ref{fig.1}(d)), an additional top gate of thickness 1 \AA\, was introduced to the device between the top part of the Source/Drain(S/D) electrodes connected to the top of an additional dielectric slab of thickness ($d_{t}$) = 4 \AA\,. The spatial separation between the molecule and both the dielectric layers is $\sim$ 1 \AA. Gold was chosen as the generic electrode material to ensure minimum contact resistance and good electrical conductivity at this temperature of operation. The equivalent capacitance network for the single and double gated devices were illustrated in Fig.~\ref{fig.1}(c) and Fig.~\ref{fig.1}(e) respectively which describes the nature of electrostatic coupling between the molecule and different electrodes. Details of the capacitance estimation methods were described at a later stage of this article.

The charging energies of this system was estimated by performing Ab-initio calculations based on Stokbro's method \cite{Stokbro2010} using the commercial Atomistic Toolkit package (ATK 11.2.3) made by the Quantum Wise Division \cite{ATK}. This self-consistent calculations were performed using a nonspin-polarised DFT framework under the Local density approximation (LDA) where the wave functions were expanded in double-$\zeta$ polarised (DZP) basis set.  Considering the metal electrodes as equipotential surfaces, Neumann boundary conditions were used while solving the Poisson's equation assuming the perpendicular component of the electric field stays at zero at the interfaces, details of which could be found here \cite{Stokbro2010}. The island-S/D electrode separation was chosen to be large enough so that calculations can be performed in the weak coupling limit as it is possible to estimate the molecular energy states accurately when incoming electron gets sufficiently enough time to stabilise on the molecule, hence carrying no information about its initial or final states. In such a device, electronic transport is only possible when an electron can be moved from the highest occupied molecular orbital (HOMO) level to the lowest occupied molecular orbital (LUMO) level and the minimum energy needed for this is called as the additional energy ($E_{a}$) which can be expressed by,
\begin{equation}
 E_{a} = (E_{n-1}-E_{n}) -  (E_{n}-E_{n+1}) =  (E_{n+1}-E_{n-1} - 2E_{n})
 \label{eqn.1}
\end{equation}
where $n$ is the number of electrons in the neutral state of the molecule.

\section{Results and Discussion}

\begin{figure}
\begin{center}
\includegraphics[width=8.5cm]{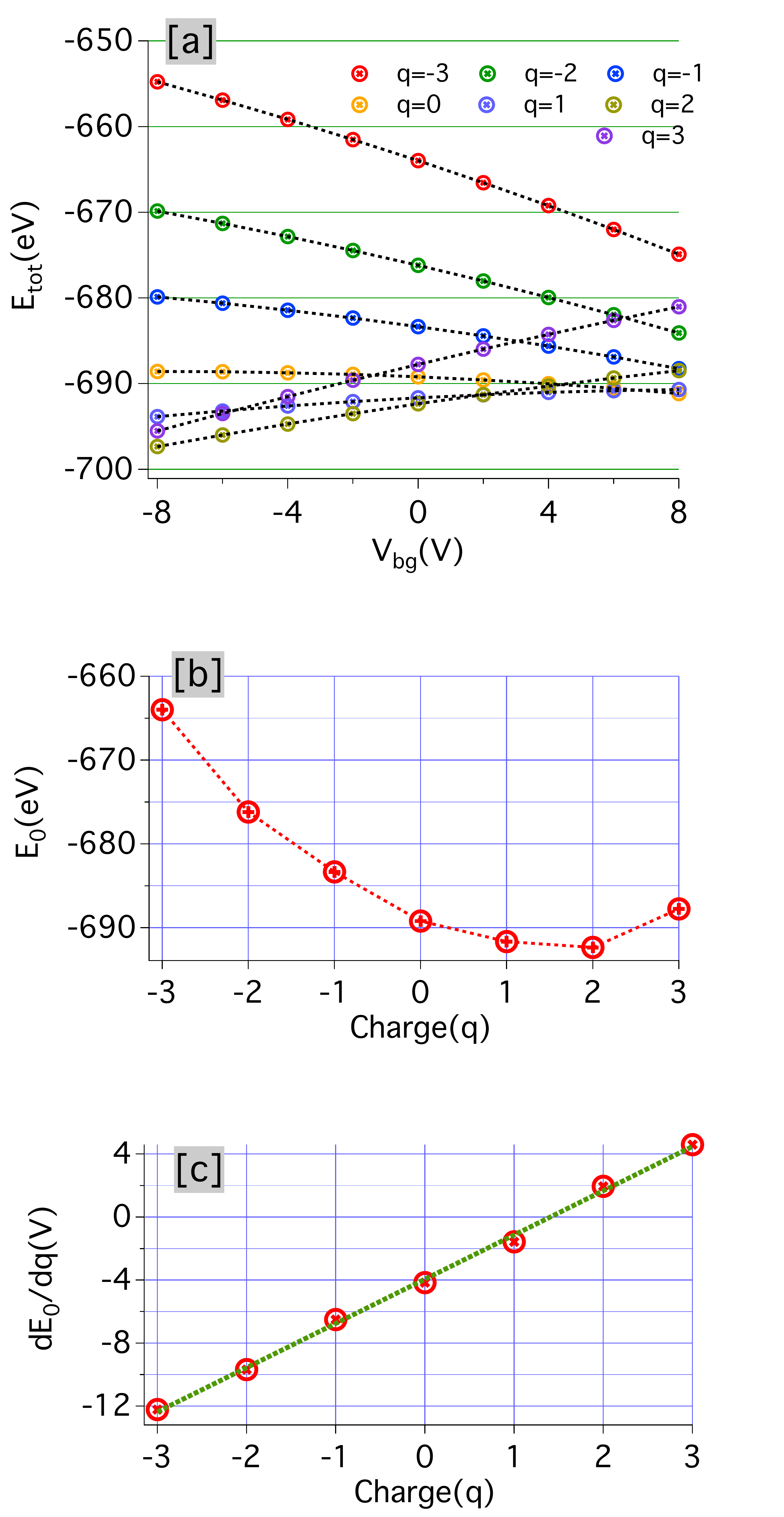}
\caption{{\small (a) Total energy of the molecule in the Double gated SMT as function of V$_{bg}$ for V$_{tg}$ = -8 V for different charge states of the molecule. The points are the calculated values and the dotted lines represent the fitted values using Eqn.~\ref{eqn.2}, (b) $E_{0}$ estimated from the fits above as a function of the charge state $q$ at V$_{tg}$ = -8 V, (c) Differential of the zero-term energy ($dE_{0}/dq$) as function of $q$, showing a linear dependence. The `green' line is the fit to the data points in `red'.}}
\label{fig.2}
\end{center}
\end{figure}

In Fig.~\ref{fig.2}(a), the total energy of the molecule in the SMT environment has been estimated for different charge states ($q$) as function of the backgate voltage ($V_{bg}$) for a fixed top-gate voltage $V_{tg}$ = -8V. A part of the total energy ($E_{tot}$) is contributed by the metallic reservoir potential (= $qW$) where $W$ is the work function of the metal electrode, which for the present case of Au electrode is 5.28 eV. For all the charge states, $E_{tot}$ showed monotonic dependence with the gate excitation. For the positive charge states, the total energy gets reduced at negative gate voltages and the situation is opposite for negative charge states where positive gate voltage reduces $E_{tot}$. This happens due to the stabilisation of the positive charge states at $V_{bg}<0$ and vice-versa for the negative charge states at $V_{bg}>0$ which can be explained from the relative orientation of the molecular energy levels for different backgate voltages. Under the application of a $V_{bg}>0$, the LUMO level goes below the Fermi level ($E_{F}$) of the electrode and this allows the transfer of an additional electron to the molecule and it becomes more negatively charged. However for $V_{bg}<0$, the HOMO level gets shifted towards a higher energy level and in this case, an electron moves from the molecule to the electrode making it more positively charged.

For investigating further dependence of $E_{tot}$ with $V_{bg}$, the energy can be analytically fitted by a $2^{nd}$ order polynomial as,
\begin{equation}
E_{tot}(q) = E_{0}(q) + \alpha qV_{bg} +\beta (eV_{bg})^{2}
\label{eqn.2}
\end{equation}
where coefficients $\alpha$ and $\beta$ are the fitting parameters which for the present case were estimated to be $\alpha$ = 0.224 and $\beta$= 0.01 eV$^{-1}$. The $2^{nd}$ term in the Eqn.~\ref{eqn.2} is proportional to $q$ which is due to the strong coupling of the backgate electrode to the molecule and the value of $\alpha$ is a measure of the strength of coupling between them. The final term in Eqn.~\ref{eqn.2} indicates the influence of the electrical polarisation of the molecule and is independent of the charge states. Since the molecule stays flat parallel to the dielectric surface and hence the difference of the electric field experienced by different atoms is minimal and this term has a smaller contribution compared to the $2^{nd}$ term as supported by $\beta < \alpha$. The values of $\alpha$ and $\beta$ stay almost constant for different charge states and the dependence of $E_{0}$ with the charge states was illustrated in Fig.~\ref{fig.2}(b). In the absence of any $V_{bg}$, $E_{0}$ decreases with an increase in $q$ initially for $q < 0$ as removing an electron at $V_{tg} < 0$ is energetically more expensive comparerd to the situation when $q \ge 0$. However, the rate of this change in $E_{0}$, $dE_{0}/dq$ which can be estimated from the slope as plotted in Fig.~\ref{fig.2}(c) is linear in $q$. The linear slope indicates that only in the presence of a $V_{tg}$, adding or removing a charge from the molecule at a specific state will have identically opposite influence on the zero-term in energy.

\begin{figure*}
\begin{center}
\includegraphics[width=17cm]{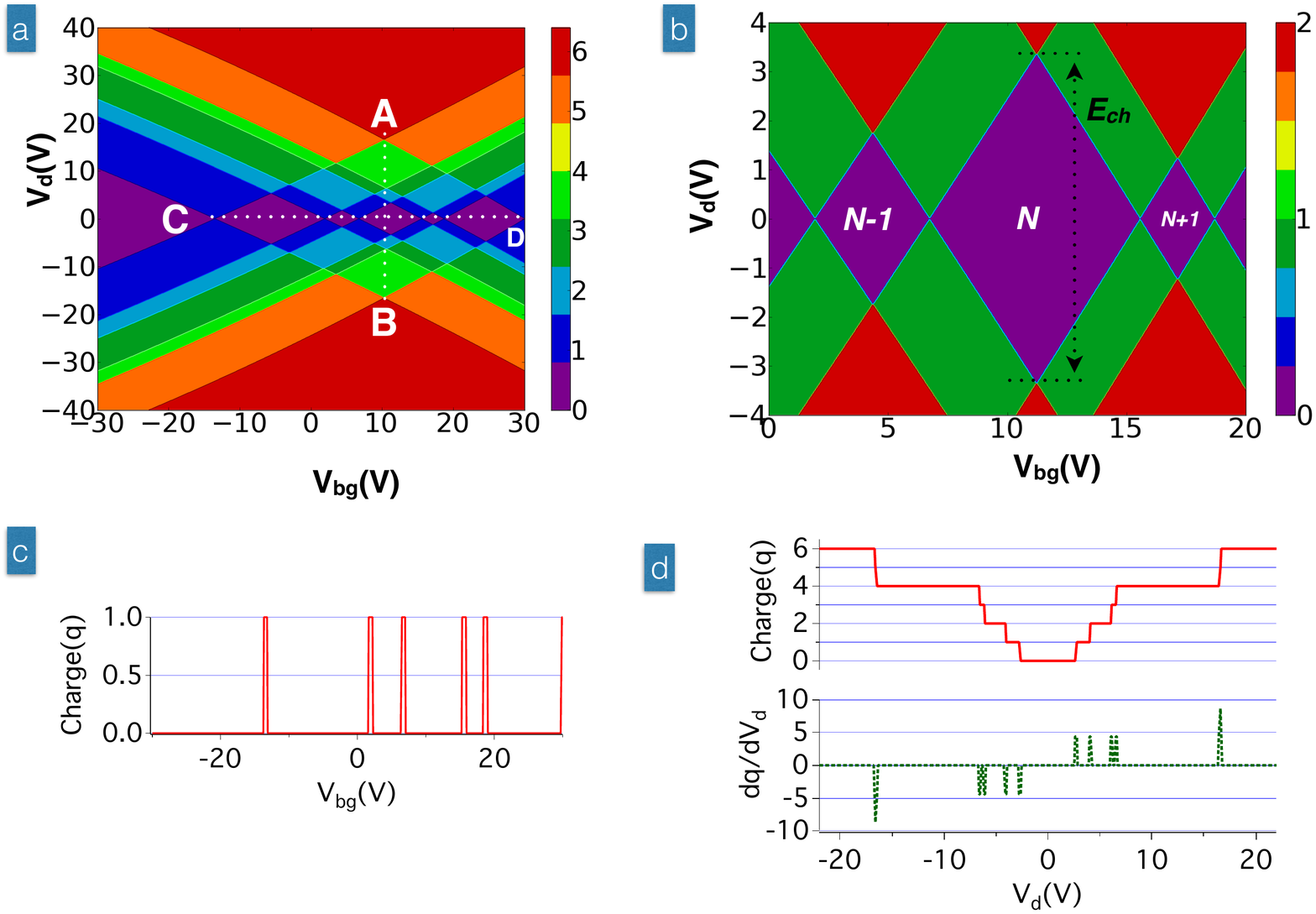}
\caption{{\small (a) Charge stability diagram estimated for the double gated SMT device for V$_{tg}$ = -8 V. (b) Zoomed version of central region of the large diamond in Fig.~\ref{fig.3}(a) to highlight the 3 central diamonds used for estimating charging energy in the ground state. (c) Line scan taken along `CD' and (d) along `AB' as marked in Fig.~\ref{fig.3}(a), (bottom) Differential charge ($dq/dV_{g}$) as function of V$_{d}$}.}
\label{fig.3}
\end{center}
\end{figure*}

For investigating further dependence of the total energy on $V_{bg}$ and the conduction behaviour, the charge stability diagrams (also called as Coulomb Diamond plot) have been plotted in Fig.~\ref{fig.3}. As the SMT operates in the weak coupling regime, electron transport between the S/D electrodes is only possible when the molecular energy levels are accessible within the applied bias ($V_{d}$) window. Details of this conduction regime could be found from the charge stability diagram as illustrated in Fig.~\ref{fig.3}(a) which was calculated for $V_{tg}$ = -8V. The colourbar on the right represents the number of available energy levels for conduction for different $V_{d}$ and $V_{bg}$. Experimentally the charge stability diagram is measured by measuring the source-drain current and the $z$-axis is represented by the current, conductivity or differential conductance. The diamonds indicate the accessible regions of conduction to separate the non-conducting and conducting regions for different values of $V_{d}$ and $V_{bg}$. In Fig.~\ref{fig.3}(a), the central region of the large diamond surrounded by points `A', `B', `C', `D' marks the key region of conduction within which sequential tunnelling between different charge states occurs in this device. In Fig.~\ref{fig.3}(b), this region was enlarged to describe the behaviour of the ground state excitation of the molecule. The central diamond in Fig.~\ref{fig.3}(b) corresponds to the neutral charge state of the molecule in its ground state with a hypothetical $N$ electrons in it. Within each of these diamonds, the population does not change on the molecule, however, moving from one of these regions to its left or right will result in changing the population.

The charging energy of the ground state can be estimated from the height of the central diamond (marked by dotted arrow in Fig.~\ref{fig.3}(b)) which is $6.628$ eV and the height of its `left' and `right' diamonds are $3.428$ eV and $2.4$ eV respectively which refers to the `cationic' and the `anionic' states of the molecule respectively. The additional energy needed for creating the `cationic' and the `anionic' states can also be estimated from Eqn.~\ref{eqn.1}. The smaller value of the charging energy of these two excited states can be accounted for their lower occupation in their frontier orbitals indicating that adding/removing an electron from the excited states are much energetically favourable than the neutral state. The slight difference between the charging energies of the two excited states are due to the differences in their HOMO-LUMO gaps which contributes a part in addition to the Coulomb energy.

From the Charge stability diagram, further information about the detailed electrostatics of the system could be obtained about the SMT. The values of the junction capacitances as indicated in Fig.~\ref{fig.1}(c),(e) could be obtained from the `ABCD' region of the diamond as indicated in Fig.~\ref{fig.3}(a) and their estimated values were listed in Table.~\ref{tab.1}. The Source and Drain capacitances i.e. $C_{s}$ and $C_{d}$ respectively were estimated from the slopes of the lines `AD' and `AC'. The slope of the line `AD' is $\sim -C_{dot-bg}/C_{d}$ and the slope of `AC' = $C_{dot-bg}/(C_{s} + C_{d})$ where $C_{dot-bg}$ is the capacitance between the `dot' and the `backgate'. The gate capacitances were estimated analytically by considering a planar approximation of the gate electrodes in a parallel plate geometry which can be expressed by : $C_{dot-bg} = \varepsilon_{0}(1+\varepsilon_{r})A_{bg}/d_{b}$ and $C_{dot-tg} = \varepsilon_{0}(1+\varepsilon_{r})A_{tg}/d_{t}$ where $\varepsilon_{r}$ is the relative permeability of the dielectric layer, $C_{dot-tg}$ is the capacitance between the `dot' and the `topgate', $d_{b}$ and $d_{t}$ are the thicknesses of the `bottom' and `top' dielectric layers respectively and $A_{bg}$, $A_{tg}$ are the areas of contacts between the `bottom' and `top' gates with their respective dielectric layers in contacts.

In Fig.~\ref{fig.3}(c), a horizontal linescan was taken along the dotted line `CD' as represented in Fig.~\ref{fig.3}(a). The appearance of sharp peaks represent the points when the conduction window changes form a certain diamond to the neighbouring one and hence the change of the `dot' population by 1 between each of the neighbouring diamonds, from left to right for increasing $V_{bg}$. At low bias voltage ($V_{d}$), the excitations only occur from the ground state to the first excited state which is why the peaks arise at values $q = 1$ for all values of $V_{bg}$. Another vertical line scan taken along `AB' (as illustrated in Fig.~\ref{fig.3}(d)) showed occurrence of periodic plateaus symmetrically placed on both sides of $V_{d} = 0$ V.  The step like structure arises only at finite values of $q$ as a result of charging between different charge states and the charging energies can be estimated from the width of the respective steps. In the Coulomb Blockade regime, the minimum occours at $V_{d}$ = 0 V in the absence of any excitation and subsequent enhancement of $V_{d}$ results in moving to the higher excited states. The occurrence of the plateaus can be understood in details from the differential ($dq/dV_{d}$) charge plot which is similar to the behaviour observed from the differential conductance (dI/dV) measured in an experiment as illustrated in the bottom panel of Fig.~\ref{fig.3}(d). The peak positions observed in Fig.~\ref{fig.3}(d) [bottom panel] represent the values of $V_{d}$ at which charging to an excited state occurs in the absence of continuous transport when the tunnelling rate ($\Gamma$) is very low. The separation between two neighbouring peaks in Fig.~\ref{fig.3}(d) [bottom panel] represents the charging energy of the molecule between subsequent excited states for a given $V_{bg}$ which is not uniform between all the charge states.

To find the systematic dependence of the $V_{tg}$ in such devices, the charging energy ($E_{ch}$) for the neutral case in the ground state of the molecule was compared for different values of $V_{tg}$. The minimum in $E_{ch}$ occours at $V_{tg}$ = -8V and it increases with an increase in the $V_{tg}$. Reduction of the $E_{ch}$ occurs when the molecule is bought from its gas phase to a SMT environment due to the polarisation of its charge states in the presence of metallic electrodes. Under the addition a top electrode, this effect gets enhanced and the non-linear dependence of $E_{ch}-V_{bg}$ indicates the influence of additional polarisation in the presence of the top electrode and a positive $V_{tg}$ induces additional polarisation to the molecule in its ground state.

\begin{figure*}
\begin{center}
\includegraphics[width=17cm]{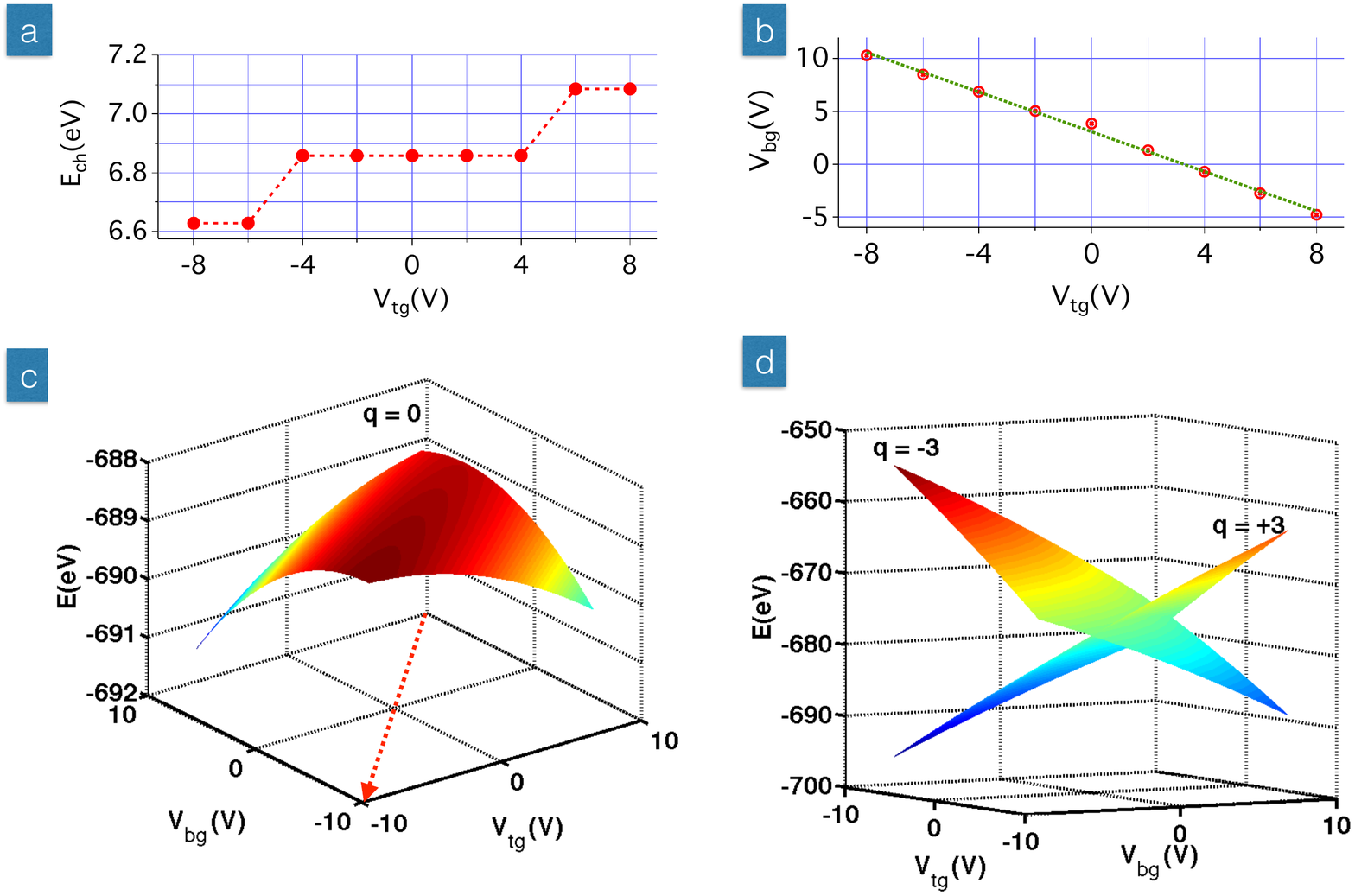}
\caption{{\small (a) Charging energy in the neutral case of the molecule in the ground state (as estimated in Fig.~\ref{fig.3}(b)) as function of V$_{tg}$, (b) Relative location of V$_{bg}$ as function of V$_{tg}$ as estimated from the coordinate of `B' of the diamond in Fig.~\ref{fig.3}(a) [see text for details], (c) Energy surface as function of  V$_{tg}$ and V$_{bg}$ for the $q$ = 0 state, The `red' arrow is the line along which two gate voltages stay at identical values,  (c) Energy surface as function of  V$_{tg}$ and V$_{bg}$ for $q$ = +3 and $q$ = -3 charge states.}}
\label{fig.4}
\end{center}
\end{figure*}

It is to be noted that the vertices of the large diamond enclosed by `ABCD' (see Fig.~\ref{fig.3}(a)) does not stay at the same position with a change of $V_{tg}$. The position of the line `AB' can be referenced to investigate this dependence. Since the $V_{bg}$ stays the same for both `A' and `B', the $x-$coordinate which is the value of $V_{bg}$, can be considered as a reference point.  In Fig.~\ref{fig.4}(b), such values of $V_{bg}$ from the line `AB'  was plotted as function of $V_{tg}$ which indicates an almost linear dependence between them. In Fig.~\ref{fig.4}(b), the dependence between the two gate voltages was plotted to estimate the nature of coupling between them through the molecule. The slope of this line is $1.062$, which is the same as the ratio of the separation between the two gates from the molecule i.e $(d_{t}+1)/(d_{b}+1)$ = 1.063. The slope $\sim$ 1 of the line indicates the almost identical electrostatic coupling of the gate electrodes to the molecule.

The electrostatic energy landscape was plotted as a 3D surface as functions of $V_{bg}$ and $V_{tg}$ in Fig.~\ref{fig.4}(c),(d). For $q=0$, the surface is almost symmetrically curved around the red dotted line as illustrated in Fig.~\ref{fig.4}(c). When the molecule is in a neutral charge state, maximum of the energy occurs due to the polarisation of the molecule when both the gates are at equal values (`red' dotted line in Fig.~\ref{fig.4}(c)) and when they are of opposite signs, total energy gets reduced as one of two gate voltages increases the polarisation while the other counter balances it. For this reason, the minimum of the energy can be found at the two extremes ($V_{tg}$ =+8V, $V_{bg}$ =-8V) and ($V_{tg}$ =-8V, $V_{bg}$ =+8V) for $q = 0$. For higher charge states, the situation changes significantly which can be seen from Fig.~\ref{fig.4}(d). For $q = -3$, the maximum in energy occurs at ($V_{tg}$ = -8V, $V_{bg}$ = -8V) and minimum on the other end at ($V_{tg}$ = +8V,$V_{bg}$ = +8V) when the charge state is more stabilised. For $q=+3$, the situation is opposite which has minimum at $V_{tg}$ = -8V, $V_{bg}$ = -8V and maximum at $V_{tg}$ = +8V, $V_{bg}$ = +8V. The nature of the energy surfaces is different for different charge states and starting from any point of any of these landscapes, energy sharply changes with a change of $V_{bg}$ and $V_{tg}$ which makes it sensitive for detection of an incoming charge state. The steep slope of the energy surfaces indicates that the energy state of the SMT is highly sensitive to a small change of the gate voltages and this property can be used as a sensitive charge detection technique. Due to the sequential nature of transport, the change of energy can be translated directly to the change of current flowing through the device ($I_{ds}$) which in the case of a real experimental device is expected to demonstrate an identical sensitivity. The difference in the nature of the energy surfaces for different charge states thus allows to detect an incoming charge state uniquely in such a device and the addition of a top gate proves to be extremely useful for investigating such behaviour.

\begin{table}
\begin{tabular}{c | c |c | c | c}
\hline  Capacitance & C$_{dot-tg}$(aF) & C$_{dot-bg}$(aF) & C$_{s}$(aF) & C$_{d}$(aF) \\
\hline  Values & 2.045 & 6.002 & 7.0325 & 7.0325\\
\hline
\end{tabular}
\caption{Junction capacitances for single and double gated devices}
\label{tab.1}
\end{table}%

\section{Conclusion}
In this work, we have performed computational investigation of a SMT with an 1,3-Cyclobutadiene molecule working as the `quantum dot'. Unlike the conventional case of a single gated device, we have found better control of the electrostatics and performance by adding a top gate electrode. The performance of such a device was investigated by calculating the energy of the molecule for different gate voltages which were used later to construct the charge stability diagram to understand the detailed nature of conduction. The molecular energy states are highly sensitive to the gate voltages that changes significantly for different charge states and this unique behaviour can be utilised for using it as a sensitive organic charge detector. These kinds of single molecular devices could be the ideal candidates for future generation nano-electronic devices as charge sensors for faster operational speeds and portability. Unlike the SET devices, these SMT's are operational at room temperature which is a genuine advantage of these molecular devices for operational usefulness and industrial applications.

\acknowledgments
SR acknowledges the support provided through a Research Excellence Grant (REG) within the Joint Research Project ``Qu-Ampere'' (SIB07) supported by the European Metrology Research Programme (EMRP). The EMRP is jointly funded by the EMRP participating countries within EURAMET and the European Union.

% ===================================================================== % %  Bibliography
% ===================================================================== %

\bibliographystyle{unsrt}

\begin{thebibliography}{11}

\bibitem{Devort1992}
M. H. Devoret, D. Esteve, and C. Urbina,
\newblock {\em Nature}, {\bf 360}, 543 (1992).


\bibitem{Delsing1998}
R. J. Schoelkopf, P. Wahlgren, A. A. Kozhevnikov, P. Delsing, and D. E. Prober,
\newblock {\em Science}, {\bf 280}, 1238 (1998).


\bibitem{Berman1997}
David Berman and Nikolai B. Zhitenev, and Raymond C. Ashoori and Henry I. Smith and Michael R. Melloch,
\newblock {\em Journal of Vacuum Science \& Technology B}, {\bf 15}, 6 (1997).


\bibitem{ATK}
Atomistix Tool Kit,
\newblock {\em Quantum Wise (http://quantumwise.com/)}.


\bibitem{Reed1997}
M. A. Reed, C. Zhou, C. J. Muller, T. P. Burgin, and J. M. Tour,
\newblock {\em Science}, {\bf 278}, 252 (1997).


\bibitem{Heath2003}
J. R. Heath and M. A. Ratner,
\newblock {\em Physics Today}, 43-49 (2003).


\bibitem{Dekker1998}
S. J. Tans, A. R. M. Verschueren, and C. Dekker,
\newblock {\em Nature}, {\bf 393}, 49 (1998).


\bibitem{Dekker2001}
H. W. C. Postma, T. Teepen, Z. Yao, M. Grifoni, and C. Dekker, 
\newblock {\em Science}, {\bf 293}, 76 (2001).


\bibitem{Park2000}
Hongkun Park and Jiwoong Park and Andrew K. L. Lim and Erik H. Anderson and A. Paul Alivisatos and Paul L. McEuen, 
\newblock {\em Nature}, {\bf 407}, 57-60 (2000).


\bibitem{Chae2006}
Dong-Hun Chae and John F. Berry and Suyong Jung and F. Albert Cotton and Carlos A. Murillo and Zhen Yao, 
\newblock {\em Nano Letters}, {\bf 6}, 2 (2006).


\bibitem{Ray2014b}
S. J. Ray, 
\newblock {\em Journal of Applied Physics}, {\bf 116}, 154302  (2014).



\bibitem{Stokbro2010}
K. Stokbro, 
\newblock {\em Journal of Physical Chemistry C}, {\bf 114}, 20461 (2010).


\bibitem{Parashar2012}
S. Parashar, P. Srivastava, and M. Pattanaik, 
\newblock {\em Applied Nanoscience}, {\bf 2}, 385 (2012).


\bibitem{Dekker2001Science}
A. Bachtold, P. Hadley, T. Nakanishi, and C. Dekker,
\newblock {\em Science}, {\bf 294}, 1317 (2001).


\bibitem{Kasper2009}
K. Moth-Poulsen and T. Bj$\o$rrnholm,
\newblock {\em Nature Nanotechnology}, {\bf 4}, 551 (2009).


\bibitem{Brandbyge2002}
M. Brandbyge, J.-L. Mozos, P.Ordej\'on, J. Taylor, and K. Stokbro, 
\newblock {\em Phys. Rev. B}, {\bf 65}, 165401 (2002).


\bibitem{Kaasbjerg2008}
K. Kaasbjerg and K. Flensberg, 
\newblock {\em Nano Letters}, {\bf 8}, 3809 (2008).


\bibitem{Ventra2000}
M. Di Ventra, S. T. Pantelides, and N. D. Lang,
\newblock {\em Phys. Rev. Lett.}, {\bf 84}, 979 (2000).


\bibitem{Guo2012}
Y.-D. Guo, X.-H. Yan, and Y. Xiao, 
\newblock {\em The Journal of Physical Chemistry C}, {\bf 116}, 21609 (2012).


\bibitem{Kubatkin2003}
Sergey Kubatkin, Andrey Danilov, Mattias Hjort, Jerome Cornil, Jean-Luc Bredas, Nicolai Stuhr-Hansen, Per Hedegard and Thomas Bj$\o$rrnholm, 
\newblock {\em Nature}, {\bf 425}, 698-701 (2003).


\bibitem{Osorio2007}
Edgar A.  Osorio, Kevin O'Neill, Maarten Wegewijs, Nicolai Stuhr-Hansen, Jens Paaske, Thomas Bj$\o$rnholm and Herre S. J Van der Zant, 
\newblock {\em Nano Letters}, {\bf 7}, 11 (2007).

\end{thebibliography}

% ===================================================================== % %  Figures
% ===================================================================== %

\newpage 

\end{document}